\documentclass[a4paper]{aa}

\usepackage{psfig,times}
\psfigurepath{/:/z/rue/plots/allessioplot:/z/rue/plots/kuekerplot:/z/rue/plots/ritterplot}
\def\ea{et\thinspace al.\,}

\def\''{\lq\lq}

\def\gsim{\lower.4ex\hbox{$\;\buildrel >\over{\scriptstyle\sim}\;$}}
\def\lsim{\lower.4ex\hbox{$\;\buildrel <\over{\scriptstyle\sim}\;$}}
\def\~  {$\sim$}

\def\newpage{\vfill\eject}

\def\qq{\qquad\qquad}
\def\qqq{\qquad\qquad\qquad}

\def\apj{{ApJ}}
\def\apjs{{ Ap. J. Suppl.}}
\def\apjl{{ Ap. J. Letters}}

\def\aa{{A\&A}}

\def\sp{{Solar Phys.}}
\def\gafd{{GAFD}}

\def\bib{\item{}}
\renewcommand{\vec}[1]{\mbox{\boldmath$#1$}}
\def\R{R\"udiger}

\def\L{$\Lambda$}
\def\Om{{\it \Omega}}
\def\nT{$\nu_{\rm T}$}     
\def\eT{$\eta_{\rm T}$}  
\input epsf

\begin{document}

%--------HEADER------------------------------------------------------------

\title{Differential rotation and meridional flow in the solar supergranulation 
layer: Measuring the eddy viscosity}
\author{G. R\"udiger\inst{1}, M. K\"uker\inst{1} \and K.L. Chan\inst{2}}
\offprints{G.~R\"udiger}
\institute{Astrophysikalisches Institut Potsdam,  An der Sternwarte 16, 
D-14482 Potsdam, Germany
\and
Department of Mathematics, The Hong Kong University of Science and Technology, 
Clear Water Bay, Hong Kong, China}
\date{\today}
\abstract{
We measure the eddy viscosity in the outermost layers of the solar convection 
zone by comparing the rotation law computed with the Reynolds stress resulting 
from f-plane simulations of the angular momentum transport in rotating 
convection with the observed differential rotation pattern. The simulations 
lead to a {\em negative } vertical  and a {\em  positive } horizontal angular 
momentum transport. The consequence is a subrotation of the outermost layers, 
as it is indeed indicated both by helioseismology and the observed rotation 
rates of sunspots. In order to reproduce the observed gradient of the rotation 
rate a value of about $1.5 \times 10^{13}$ cm$^2$/s for the eddy viscosity is 
necessary. Comparison with the magnetic eddy diffusivity derived from the 
sunspot decay yields a surprisingly large magnetic Prandtl number of 150 for 
the supergranulation layer. The negative gradient of the rotation rate also 
drives a surface meridional flow towards the poles, in agreement with the 
results from Doppler measurements. The successful  reproduction of the {\em 
abnormally positive} horizontal cross correlation (on the northern hemisphere) 
observed for bipolar groups then provides an independent test for the resulting 
eddy viscosity. 
\keywords{Turbulence - Sun: rotation}
}
\authorrunning{G. R\"udiger,  M. K\"uker \& K.L. Chan}
\titlerunning{Differential rotation and meridional flow in the solar 
supergranulation layer}
\maketitle

%--------SECTION1------------------------------------------------------------
%%%%%%%%%%%%%%%%%%%%%%%%%%%%%%%%%%%%%%%%%%%%%%%%%%%%%%%%%%%%%
\section{Introduction and observations}
\begin{figure}[ht]
\center
\mbox{
\psfig{figure=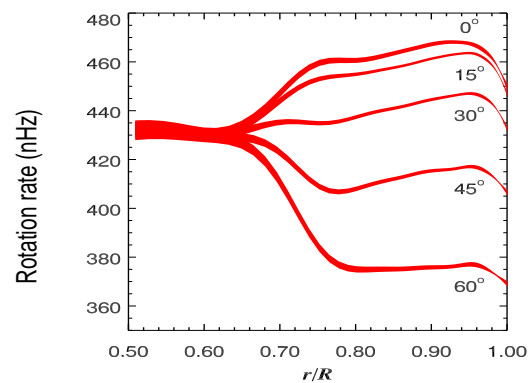,width=7cm,height=5cm}
}
\label{f1}
\caption{The internal rotation of the Sun as found by helioseismology. 
Image: NSF's National Solar Observatory
} 
\end{figure}

%%%%%%%%%%%%%%%%%%%%%%%%%%%%%%%%%%%%%%%%%%%%%%%%%%%%%%%%%%%%%%
%
Over the last years, while observing the solar oscillations on longer
timescales and with higher precision, it has become evident that these
oscillations play an important role in understanding the solar interior,
bearing more or less the only information from the deeper parts of the sun, 
which cannot be probed otherwise. Helioseismology reveals a {\em maximum} of 
the angular velocity at {\em all} latitudes rather close to the surface, as 
shown in Fig.~\ref{f1} (Howe et al.~2000). 

It  is known, on the other hand, that sunspots rotate faster than the 
solar surface plasma by about 4 \% or 80 m/s at all latitudes. Such a clearly 
verified subrotation of the outermost layer of the convection zone is easiest 
understood as a result of angular momentum conservation of fluid elements with  
purely radial motions. But in this domain of the solar convection zone the 
velocity field is dominated by horizontal motions. Fluctuating fields with 
predominantly horizontal intensity should produce superrotation rather than 
subrotation. There is, however, another strong argument for considering the 
exceptional behavior of the {\em horizontal motions} in more detail.

Ward (1965) was the first to consider the horizontal cross-correlation 
of the proper motions of sunspot groups, the faster of which tend to move 
toward the equator. He found
\begin{equation}
Q_{\theta\phi} \approx \ 0.1\  ({\rm deg/day})^2 \approx 2 \times 10^7 {\rm 
cm}^2/
{\rm s}^2
\label{0}
\end{equation}
on the northern hemisphere. More recent observations found smaller, but 
always positive values (Gilman \& Howard 1984; Nesme-Ribes \ea 1993; 
Komm \ea 1994, see an overview  by Meunier et al. 1997). This result has a 
strong implication for theory confirming the existence of the positive $H$ 
coefficient in the expression (\ref{3}) for the horizontal Reynolds stress. 
We shall demonstrate the close relation between the negative 
radial gradient of the rotation rate and the positive sign of the horizontal 
cross correlation by solving the Reynolds equation on the basis of 
new data from hydrodynamic simulations of rotating convection in the outermost 
layer of the convection zone. The computations provide a tool for measuring 
the eddy viscosity in the outer solar convection zone.
%
%%%%%%%%%%%%%%%%%%%%%%%%%%%%%%%%%%%%%%%%%%%%%%%%%%%%%%%%%%%%%%%%%%%%%%%%%%%
%
\section{The Reynolds stress}
%
%%%%%%%%%%%%%%%%%%%%%%%%%%%%%%%%%%%%%%%%%%%%%%%%%%%%%%%%%%%%%%%%%%%%%%%%%%%
%
We start with the construction of the two components of the Reynolds stress 
tensor,
\begin{equation}
Q_{ij} = \langle u_i'(\vec{x},t)u_j'(\vec{x},t)\rangle,
\label{2}
\end{equation}
which describe the angular momentum transport, i.e.
\begin{eqnarray}
Q_{r\phi} &=& - \nu_{\rm T} r {\partial \Om \over \partial r} \sin\theta + 
I\cdot V \sin\theta,\nonumber\\
Q_{\theta\phi}&=& - \nu_{\rm T} {\partial \Om \over \partial \theta} \sin\theta 
+ I\cdot H \cos\theta.
\label{3}
\end{eqnarray}
Due to the terms containing the quantity $I$, angular momentum is transported 
even in case of rigid rotation, $\Om =$ const.. This non-diffusive part of the 
stress, which only exists in case of anisotropic turbulence subject to a basic 
rotation, is  known as the $\Lambda$-effect (\R\ 1989). 
The problem of the angular velocity gradient in the outer part of the solar 
convection zone has already been discussed by Gilman \& Foukal (1979) and  
Gailitis \& \R\ (1981) with theories based on the $\Lambda$-effect. 

The stress (\ref{3}) contains three unknown parameters, which determine the 
internal rotation law: the vertical angular momentum transport $V$, the 
horizontal angular momentum transport $H$, and the eddy viscosity \nT. If the 
rotation law is known from observations and the $\Lambda$-effect is known from 
simulations, one should be able to measure the eddy viscosity. Once all three 
parameters are known we can compute a theoretical Ward profile, which can then 
be compared with the observed one as a test. 

In (\ref{3}), we have introduced the turbulence intensity,  
\begin{equation}
I= \sqrt{\langle u_r'^2\rangle \langle u_\phi'^2\rangle},
\label{4}
\end{equation}
from the simulations to normalize the $\Lambda$-effect. The functions $V$ and 
$H$ are also known from the simulations, but not the eddy viscosity, which we 
shall choose such that the observations are reproduced. With the same 
normalization for the eddy viscosity as for the $\Lambda$-effect, i.e. 
\begin{equation}
\nu_{\rm T} = \tilde \nu \  {I\over \Om},
\label{5}
\end{equation}
the Reynolds stress reads
\begin{eqnarray}
Q_{r\phi}&=& {I\over \Om} \bigg(-\tilde \nu r {\partial \Om \over 
\partial r} + V \Om\bigg) \sin\theta,\nonumber\\
Q_{\theta\phi}&=& {I\over \Om} \bigg( - \tilde \nu {\partial \Om \over 
\partial \theta} \sin\theta + H \Om \cos\theta\bigg),
\label{6}
\end{eqnarray}
with $\tilde \nu$ as the only free parameter. It is varied between 0 and 1 
in order to reproduce the observed value of 5\% for the decrease of the 
rotation rate with radius. It is obvious that large values of $\tilde \nu$ will 
produce negative (positive) cross correlations on the northern (southern) 
hemisphere, while small numbers of   $\tilde \nu$ and positive values of $H$ 
are required for the observed (opposite) behavior. 
Indeed, as we have scaled  the eddy viscosity with the rotation period rather 
than with the convective turnover time in (\ref{5}), we must expect a rather 
small value for $\tilde \nu$. As the eddy viscosity in the solar convection 
zone fixes the value of the Taylor number, the Taylor number fixes the 
amplitude of the meridional flow, and the meridional flow might fix the solar 
dynamo, it is important to derive the eddy viscosity from observations.
%
%
%%%%%%%%%%%%%%%%%%%%%%%%%%%%%%%%%%%%%%%%%%%
%
\section{The simulations}
%
%%%%%%%%%%%%%%%%%%%%%%%%%%%%%%%%%%%%%%%%%%%
%
 We obtain the functions $V$ and $H$ from the f-plane numerical
simulations of Chan (2001).  These calculations computed rotating
convection in local pieces of the spherical shell and obtained values
of the turbulence Reynolds stress at a different latitudes.  The
latitudinal coverage of the cases is dense enough for us to obtain
analytical fits of the numerical data. The strong density stratification is 
completely included in the model simulations, as is the energy transport by 
radiation (gradient of temperature) and convection (gradient of entropy). 

\begin{figure}[ht]
\mbox{
\psfig{figure=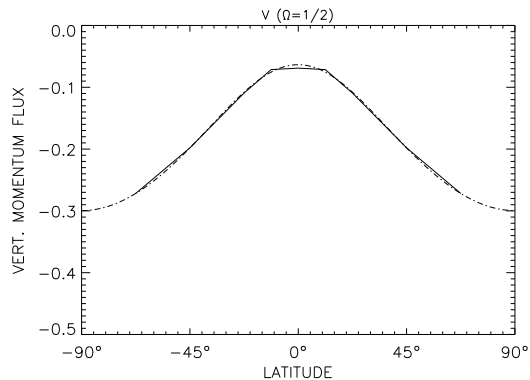,height=5.0cm,width=7cm}
}
\caption{The latitudinal profile of the vertical $\Lambda$-effect, $V$. Solid 
line: values from Chan (2001), dash-dotted line: The profile described by Eq. 
(\ref{7})} 
\label{f2}
\end{figure}

\begin{figure}[ht]
\mbox{
\psfig{figure=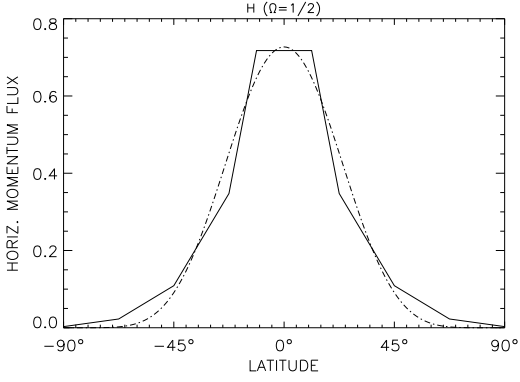,height=5.0cm,width=7cm}
}
\caption{The latitudinal profile of the horizontal $\Lambda$-effect, $H$. Solid 
line: values from Chan (2001), dash-dotted line: The profile described by Eq. 
(\ref{8})} 
\label{f3}
\end{figure}

The resulting  $\Lambda$-effect from the $\Om=1/2, F/0.25=1/8$ case, as 
tabulated in Chan (2001, Table 3), is  given in the  Figs.~\ref{f2} and 
\ref{f3} showing the data from the simulation 
vs.~the expansions of Eqs.~(\ref{7}) and (\ref{8}). The main feature of the 
vertical $\Lambda$-effect, $V$, is its negative sign (see Fig. \ref{f2})  while 
the horizontal  $\Lambda$-effect, $H$,  proves to be positive (see Fig. 
\ref{f3}).

After the theory of the $\Lambda$-effect, the vertical transport can be 
approximated with $V \propto \langle u_\phi'^2\rangle - \langle u_r'^2\rangle$ 
for slow rotation, so that predominantly vertical turbulence, $\langle 
u_r'^2\rangle > \langle u_\phi'^2\rangle$, is needed for $V$ to assume 
negative values. As Table 2 in Chan (2001) shows, this is indeed the case.

By the signs of both $H$ and $V$, these results are very promising in order 
to reproduce the negative slope of the rotation rate as well as the 
positive sign of the horizontal cross correlations. However, in contrast to 
expectations, $V$ exhibits a minimum at the equator while $H$ is 
highly  concentrated at low latitudes. 

\begin{figure}[ht]
\mbox{
\psfig{figure=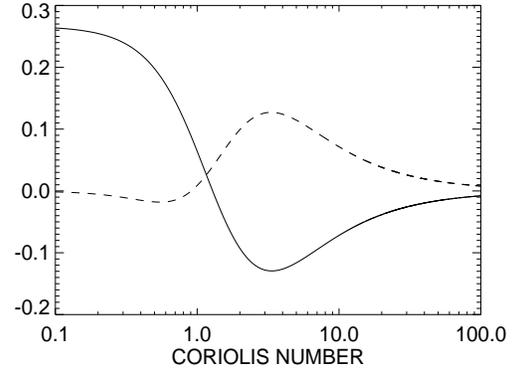,width=7.5cm,height=5.5cm}
}
\caption{
The turbulence-originated coefficients in the non-diffusive fluxes of angular 
momentum in the solar convection zone from KR93. 
Solid line: $V^{(0)}$. Dashed line: $V^{(1)}=\vec H^{(1)}$. Note that $V^{(0)}$ 
is 
positive for slow rotation 
} 
\label{lambda}
\end{figure}

As in R\"udiger (1989), the 
$\Lambda$-effect terms in 
(\ref{7}) and (\ref{8}) are written as
\begin{equation}
V= \sum_{l=0} V^{(l)} \sin^{2l}\theta, \quad H=\sum_{l=1} H^{(l)} \sin^{2l} 
\theta,
\label{8.1}
\end{equation}
so that the result of the simulations can be summarized with the series 
expansion  
\begin{eqnarray}
V&=& -0.300 + 0.187 \sin^2\theta + 0.0158\sin^4\theta +\nonumber\\
&& \qqq + 0.0337 \sin^6\theta,
\label{7}
\end{eqnarray}
and
\begin{equation}
H=0.727 \sin^6\theta,
\label{8}
\end{equation}
so that
\begin{eqnarray} 
% \label{8.2}
 V^{(0)} &=& -0.30, \qqq \ \         V^{(1)} = 0.19, \nonumber\\
 V^{(2)} &=& \ \ \ 0.016, \qqq     V^{(3)} = 0.034, \label{8.2} \\
 H^{(1)} &=& H^{(2)}= 0, \qq \ \ \ \ \ H^{(3)} = 0.73 \nonumber
\end{eqnarray}
results.
These results  are very interesting insofar as the surface effects of the
convection zone are included in the model, in contrast to the computation of 
the \L-effect for free turbulence by Kitchatinov  \& \R\ (1993, KR93). In the 
latter model the \L-effect is a function of the Coriolis number,
\begin{equation}
 \Omega^* = 4 \pi \frac{\tau_{\rm corr}}{P_{\rm rot}},
\end{equation}
where $\tau_{\rm corr}$ is the convective turnover time and $P_{\rm rot}$ the 
rotation period. The result is shown in Fig.~\ref{lambda}. For slow rotation, 
$\Omega^* \ll 1$, positive values result for $V^{(0)}$,  and  $V^{(1)}=H^{(1)}$ 
is small but negative. On the other hand, for fast rotation one finds
$V^{(0)}<0$ and $V^{(1)}=H^{(1)} \gsim 0$ similar to the numbers in Figs.
\ref{f2} and \ref{f3}. In the supergranulation layer of the solar convection
zone, however, one cannot apply the results for $\Omega^* \gg 1$. 

The values in (\ref{8.2}) differ considerably from the predictions of the 
KR93 theory indicating that the latter
is invalid for the outermost layers of the convection zone, where surface 
effects such as the presence of a boundary, strong density stratification, and 
radiation transport are essential.  
%
%%%%%%%%%%%%%%%%%%%%%%%%%%%%%%%%%%%%%%%%%%%%%%%%%%%%%%%%%%%%%%%%%%%%%
%
\section{The solution}
%
%%%%%%%%%%%%%%%%%%%%%%%%%%%%%%%%%%%%%%%%%%%%%%%%%%%%%%%%%%%%%%%%%%%%%
%
We are looking for the eddy viscosity parameter $\tilde \nu$. It must be 
positive  and it must reproduce the 5\% radial decrease of the rotation rate 
through the supergranulation layer. With the expressions (\ref{6}), we solve 
the Reynolds equation,
\begin{equation} \label{reynolds}
  \rho \left [ \frac{\partial \vec{\bar{u}}}{\partial t}
      + (\vec{\bar{u}}\cdot \nabla)
       \vec{\bar{u}} \right ] =  - \nabla p - \nabla \cdot (\rho Q)
	    + \rho \vec{g},
\end{equation}
assuming axisymmetry and under the anelastic approximation,
\begin{equation}
 \nabla \cdot (\rho \vec{\bar{u}}) = 0.
\end{equation}
In cases where the meridional flow can be neglected, the azimuthal component of 
Eq.~(\ref{reynolds}) is reduced to 
\begin{equation}
\nabla \cdot (r\sin\theta \rho \langle u'_\phi \vec{ u'}\rangle)=0.
\label{9}
\end{equation}
 Expanding the rotation rate in 
terms of Legendre functions,  
\begin{equation}
\Om= \Om_0 \sum \omega_{n-1}(x) {P_n^{(1)}(\cos \theta) \over \sin\theta},
\label{9.1}
\end{equation}
Eq. (\ref{9}) can be solved analytically for a thin layer with stress-free 
boundary conditions,
\begin{equation}
 Q_{r \phi} = 0,
\end{equation}
as described in detail by R\"udiger (1989). The boundary conditions,
\begin{equation}
\tilde\nu \omega_0' = V_0 \quad \tilde\nu \omega_2' = V_2,
\label{9.15}
\end{equation}
where $V_n$ now are the components of the function $V$ expanded in terms of 
orthogonal polynomials, lead to 
\begin{eqnarray}
V_0&=& V^{(0)} + 0.8 V^{(1)} + 0.069 V^{(2)} \simeq -0.15,\nonumber\\
V_1&\simeq& -0.02.
\label{9.18}
\end{eqnarray}
Inserting the slope $\omega_0' \simeq -1.4$ from the observed rotation law in 
Eqs. (\ref{9.15}) and  (\ref{9.18}), we obtain 
\begin{equation}
\tilde\nu \simeq 0.11
\label{9.19}
\end{equation}
for the eddy viscosity.

\begin{figure*}[ht]
\mbox{
%\hspace{-2mm}
%\psfig{figure=om_nu03.ps,height=4cm,width=2.9cm}\hfill
%\psfig{figure=om_nu01.ps,height=4cm,width=2.9cm}\hfill
%\psfig{figure=om_nu003.ps,height=4cm,width=2.9cm}
\psfig{figure=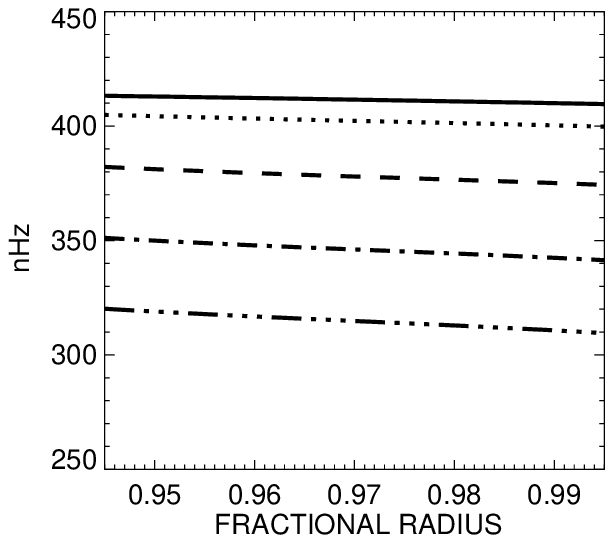,height=5.0cm,width=5.5cm}\hfill
\psfig{figure=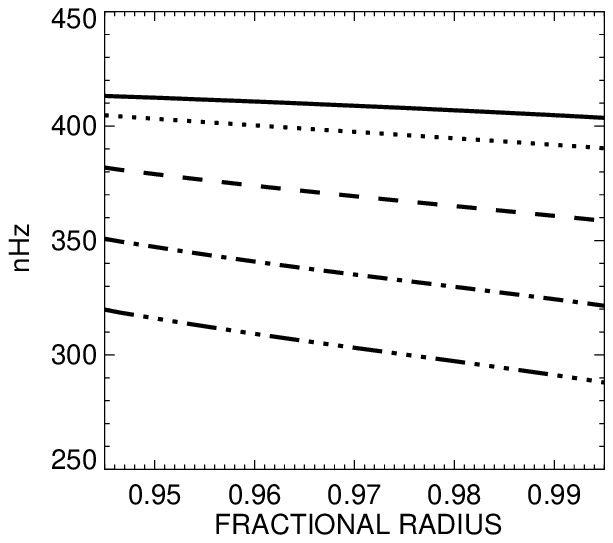,height=5.0cm,width=5.5cm}\hfill
\psfig{figure=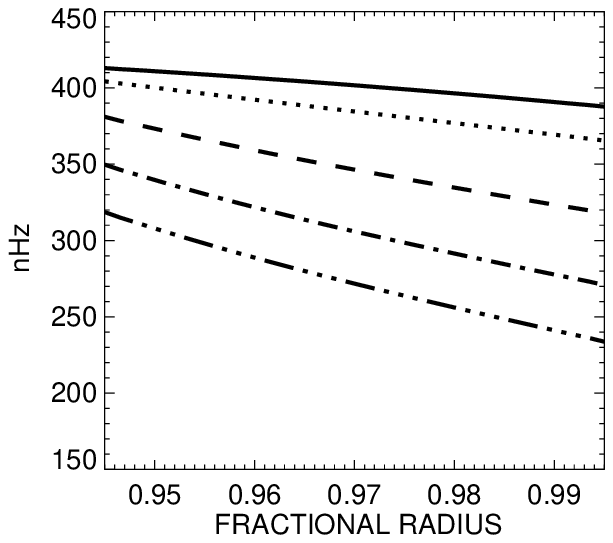,height=5.0cm,width=5.5cm}
}
\caption{
Rotation rate at the equator, 15, 30, 45, and 60 degrees latitude (from top to 
bottom) in the outermost region of the solar convection zone for $\tilde\nu 
=0.3$ (left), $\tilde\nu =0.1$ (middle), $\tilde\nu =0.03$ (right)
} 
\label{f4}
\end{figure*}

The derivation of the result (\ref{9.19}) is only valid for a very thin surface 
layer. To 
treat a layer of finite depth, we solve Eq. (\ref{reynolds}) numerically with 
$\tilde 
\nu$ as an input parameter. At the lower boundary, at $x_{\rm in}$=0.95, we 
prescribe the rotation law,
\begin{equation}
  \Om(\theta)= \Om_0 (0.7 + 0.3 \sin^2\theta),
\end{equation}
to impose the observed latitudinal shear.
The upper boundary is stress-free. The density stratification is taken from a 
standard solar model (Ahrens et al.~1992). The code used is the same as in 
K\"uker \& Stix (2001). We have made runs with and without the meridional flow 
included and found no significant difference between the results. The results 
shown in Fig. \ref{f4} were obtained with the meridional flow included. 

Indeed a negative slope of the rotation law results for {\em all latitudes}. 
Its amplitude grows with decreasing viscosity. For 
\begin{equation}
\tilde \nu = 0.1
\label{tildnu}
\end{equation}
 the results  in the Fig. \ref{f4} comply with the observations. Again the 
$\partial 
\Om/\partial r$ is negative for all latitudes.

%%%%%%%%%%%%%%%%%%%%%%%%%%%%%%%%%%%%%%%%%%%%%%%%%%%%%%%%%%%%%%%%
%
\section{Eddy viscosity and magnetic Prandtl number}
\begin{figure}[ht]
\psfig{figure=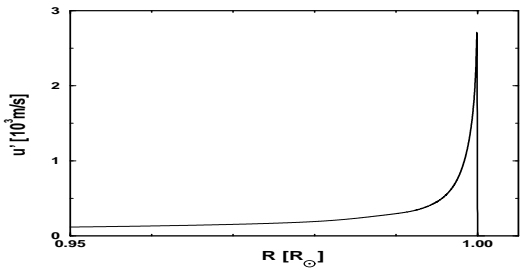,height=4cm,width=8cm}
\caption{The turbulence intensity in the solar supergranulation layer after 
Ahrens et al.~(1992)} 
\label{f8} 
\end{figure}

%
%%%%%%%%%%%%%%%%%%%%%%%%%%%%%%%%%%%%%%%%%%%%%%%%%%%%%%%%%%%%%%%%
%
As there are no observations about (say) the decay of large-scale vortices at 
the solar surface, we have no direct information about the amplitude of the 
eddy viscosity. The only way to determine it is the study of the internal 
differential rotation of the Sun. 
We have shown that for $\tilde \nu \simeq$ 0.1 the observed radial gradients of 
the outer solar rotation law   can be reproduced.  
It might be 
interesting to use this value to estimate the  eddy viscosity amplitude in the 
outermost layers of the solar convection zone. 
With a RMS value of about 200 m/s for the velocity fluctuations (see Fig. 
\ref{f8}), 
\begin{equation}
\nu_{\rm T} \simeq 1.5 \times 10^{13} \ \  {\rm cm^2/s}
\label{nut}
\end{equation}
follows from Eq.~(\ref{5}).

In stellar magnetohydrodynamics, the turbulent magnetic Prandtl number, 
Pm=\nT/\eT, is an important parameter. It is often assumed to be of order 
unity, but this  is not finally clear. The advection-dominated solar dynamo, 
e.g., requires an eddy magnetic diffusivity smaller than $10^{12}$ cm$^2$/s
(Choudhuri et al.~1995, Dikpati \& Charbonneau 1999, K\"uker et al.~2001),  
hence after (\ref{nut}) a turbulent magnetic Prandtl number greater than 15. 
The value increases to Pm $\simeq 150$ if -- as it is suggestive -- the eddy 
magnetic diffusivity, \eT, is derived from the sunspot decay ($\simeq 
10^{11}$cm$^2$/s). The dispersal of large-scale patterns in the surface 
magnetic flux, however indicates a considerably larger value of $6 \times 
10^{12}$ cm$^2$/s for the eddy magnetic diffusivity (Sheeley 1992), hence a 
turbulent magnetic Prandtl number of about 2.5.

Large values of the magnetic Prandtl number are not quite unreasonable. It is 
shown in \R\ (1989) that the turbulent Prandtl number runs with the inverse
microscopic Prandtl number, which takes the value of 0.01 for the solar plasma 
and thus would indeed lead to values of about 100 for the turbulent magnetic 
Prandtl number\footnote{The considerations in \R\ (1989) only concern the 
Prandtl number rather than the magnetic Prandtl number but the expressions are 
very similar in both cases}. 
%
%%%%%%%%%%%%%%%%%%%%%%%%%%%%%%%%%%%%%%%%%%%%%%%%%%%%%%%%%%%%%%%%%%%
\section{The Ward profile}
\begin{figure}[ht]
\center
\mbox{
\psfig{figure=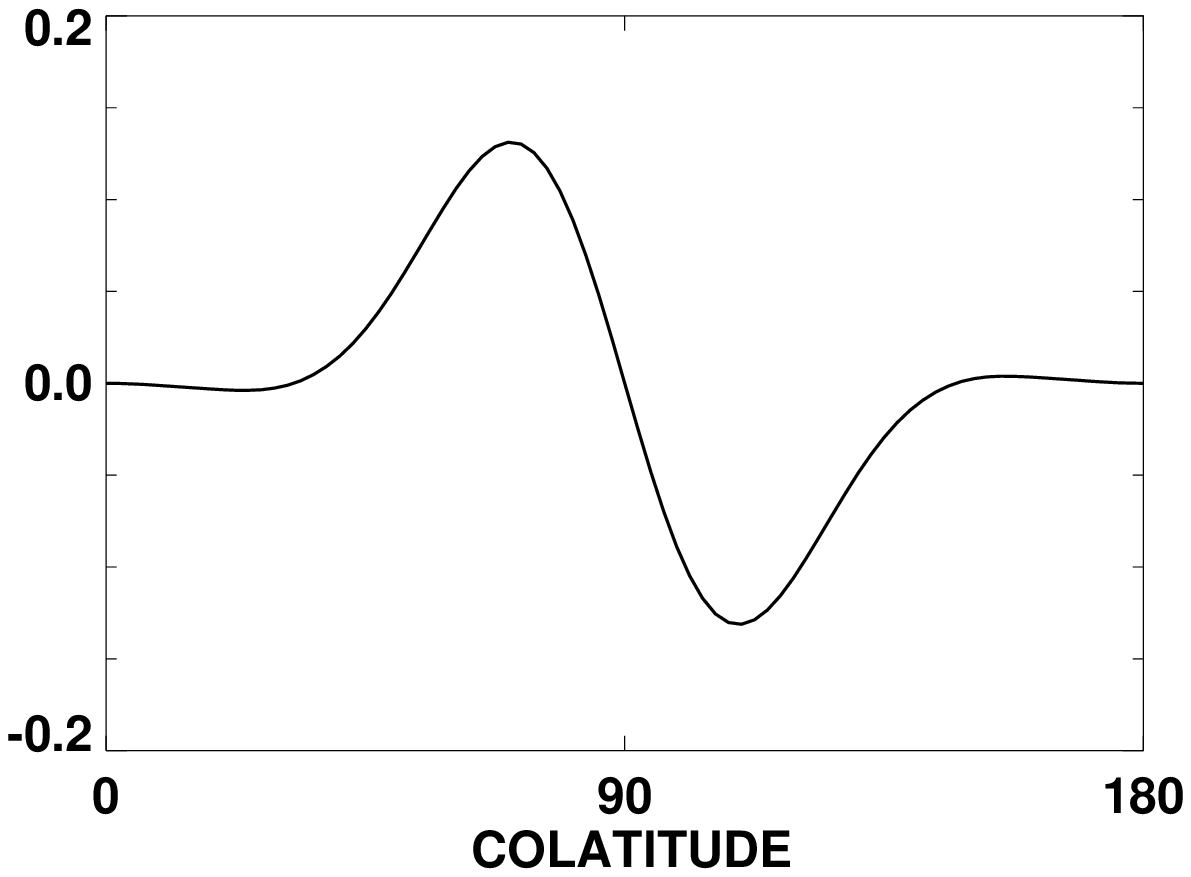,height=3.0cm,width=7cm}
}
\mbox{
\psfig{figure=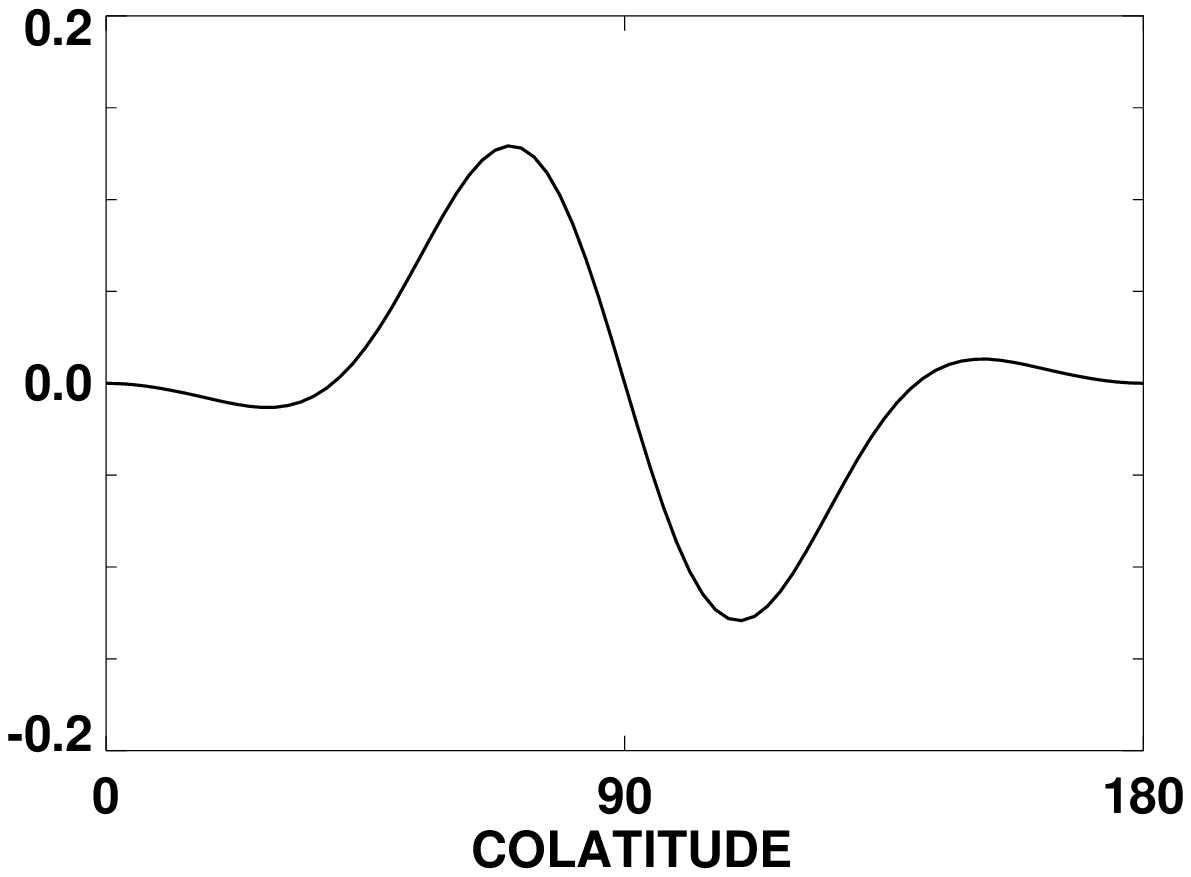,height=3.0cm,width=7cm}
}
\mbox{
\psfig{figure=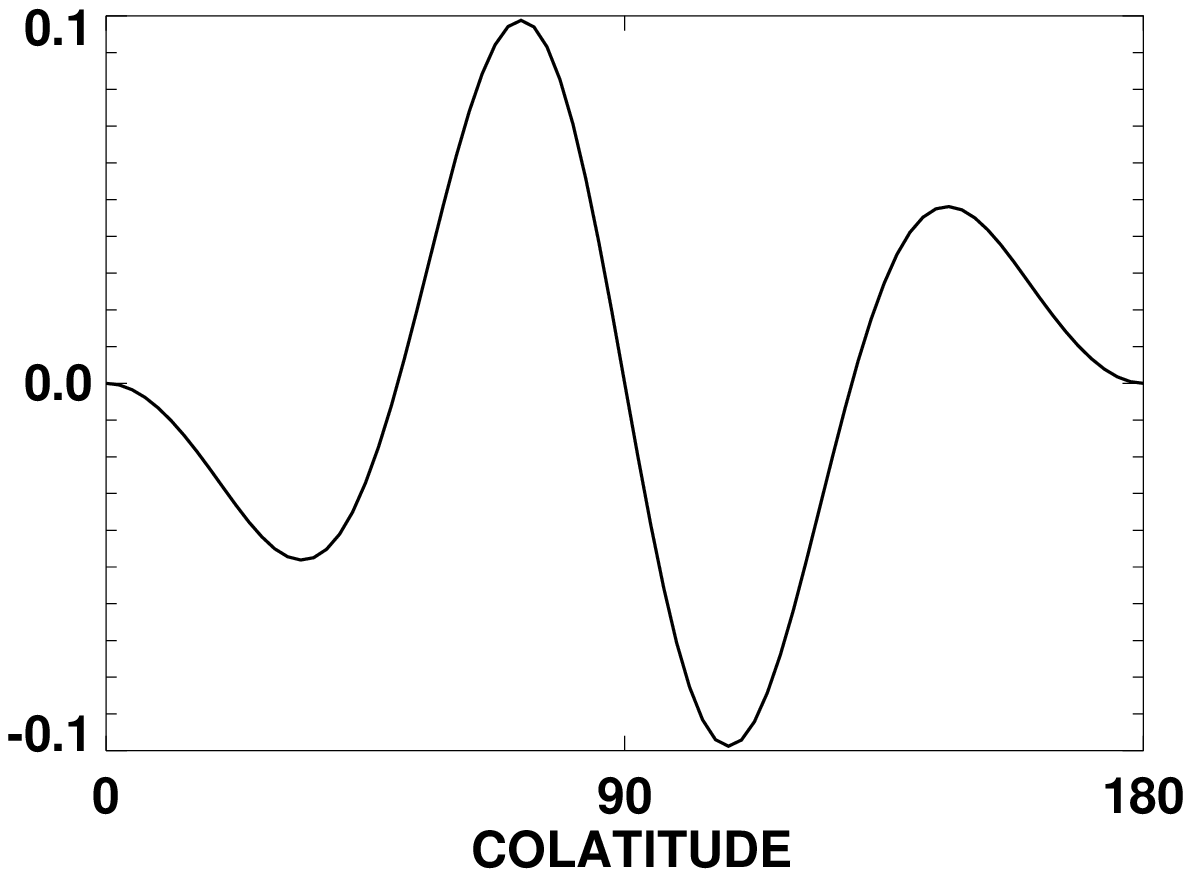,height=3.0cm,width=7cm}
}
\caption{The Ward profile W for $\tilde\nu =0.3, 0.1, 0.03$ (from top to 
bottom). The correlation is slightly { negative} in  the high  northern 
latitudes  but it is positive as observed in low northern latitudes} 
\label{f5}
\end{figure}

%%%%%%%%%%%%%%%%%%%%%%%%%%%%%%%%%%%%%%%%%%%%%%%%%%%%%%%%%%%%%%%%%%%
%
An important test is whether  such viscosity values would generate the 
positive Ward profile. This is indeed the case. At low northern latitudes the 
horizontal cross correlation is  {\em positive}, but we do not have any 
information about the higher latitudes. From the stress (\ref{6}), we find for 
the Ward profile, 
\begin{equation}
W =  - {\tilde \nu \over \Om} {\partial \Om \over 
\partial \theta} \sin\theta + H  \cos\theta.
\label{9.2}
\end{equation}
The results for various values of $\tilde \nu$ are shown in Fig.~\ref{f5}. Note 
that the covariance is indeed positive (in the northern hemisphere) at low 
latitudes, and slightly negative at high latitudes, from where we do not have 
any reliable data.

%%%%%%%%%%%%%%%%%%%%%%%%%%%%%%%%%%%%%%%%%%%%%%
%
\section{The meridional flow}
\begin{figure}[ht]
\psfig{figure=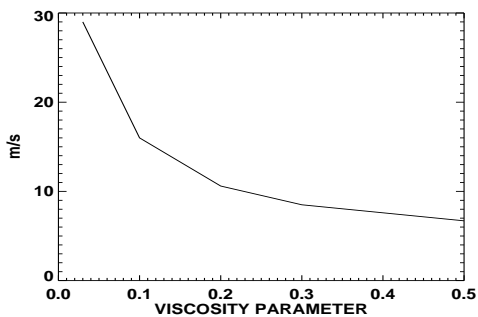,width=7.0cm,height=4.5cm}
\caption{The amplitude of the poleward directed meridional flow at the solar 
surface} 
\label{f11}
\end{figure}

Doppler measurements find a surface meridional flow that is directed polewards 
with maximum speeds between 10 and 15 m/s (Komm et al.~1993). Models of solar 
differential rotation based on the KR93 Reynolds stress usually fail to produce 
this surface flow, though a corresponding flow cell exists in the bulk of the 
convection zone. It is, however, superseeded by a second flow cell in the 
surface layers with opposite flow direction, hence producing a surface flow 
towards the equator. This superficial flow cell is driven by the positive 
radial gradient of the rotation rate, which is due to the short convective 
turnover time (and thus small Coriolis number) in the supergranulation layer. 
Since the meridional flow is mainly driven by the gradient of the rotation rate 
in $z$-direction, the effect on the flow pattern is profound. With the 
\L-effect from the Chan (2001) simulations, both the positive rotational shear 
and the additional flow cell vanish. Figure \ref{f11} shows the maximum surface 
flow speed as a function  of the viscosity parameter $\tilde{\nu}$. Contrary to 
Kitchatinov \& R\"udiger (1999) and K\"uker \& Stix (2001), the surface flow is 
directed towards the poles, and (again) for the $\tilde{\nu}=0.1$ case (which 
best reproduces the radial shear) the flow speed lies in the observed range.   
\section{Conclusions}
The theory of turbulent angular momentum transport in stellar convection zones 
based on the Second Order Correlation Approximation (SOCA) in KR93 predicts a 
positive vertical and vanishing horizontal \L-effect in the solar 
supergranulation layer, and negative vertical as well as positive horizontal 
\L-effect in the bulk of the solar convection zone. Solutions of the Reynolds 
equation with the stress tensor from KR93 therefore reproduce the observed 
variation of the rotation rate with latitude remarkably well, but lack the 
decrease of the rotation rate with increasing radius in the outermost part of 
the convection zone.  

Simulations of rotating convection in the upper part of the solar convection 
zone show a strong and positive horizontal and a negative vertical \L-effect. 
With the \L-coefficients as derived from the simulations, the solutions show a 
negative shear, $\partial \Omega/\partial r < 0$, of the observed 
amplitude when a value of $1.5 \times 10^{13} {\rm cm}^2/{\rm s}$ is chosen for 
the eddy viscosity. As there is no way to directly measure the eddy 
viscosity, this is the only method to derive its value from observations. 

As an independent test, we have computed theoretical Ward profiles. With the 
\L-effect from KR93, the horizontal cross-correlation is always negative  
because the horizontal \L-effect vanishes in the surface layer. With the large 
positive value of $H$ from the simulations, on the other hand, $Q_{\theta 
\phi}$ is always positive at low latitudes, as observed, and the amplitudes are 
in good agreement as well. 
 
In models of the solar differential rotation with positive radial \L-effect in 
the outermost layers of the convection zone, the radial shear is always 
positive in that layer, and the surface gas flow is directed towards the 
equator, both in contradiction to the observations. The negative radial 
\L-effect derived from the Chan (2001) simulations removes both these 
contradictions by maintaining a negative gradient of the rotation rate, which 
in turn drives the surface flow towards the poles.

We conclude that the KR93 expressions for the Reynolds stress are invalid in 
the outermost part of the solar convection zone. Possible reasons are the 
proximinty of the outer boundary, the increasing importance of radiative energy 
transport with decreasing depth, and the neglect of the inherent anisotropy of 
turbulent convection, which is driven by a large-scale entropy gradient rather 
than a random force.

%--------LITERATURE------------------------------------------------------------

%
\end{document}